\documentclass[twocolumn,showpacs,floatfix,superscriptaddress,amsmath,amssymb,prl]{revtex4}
\usepackage{amsfonts}
\usepackage{bbm}
\usepackage{mathrsfs}
\usepackage{txfonts}
\usepackage{amssymb}
\usepackage{graphicx}
\usepackage{hyperref}
\usepackage{ulem}
 \usepackage{overpic}
 \usepackage{psfrag}
\usepackage{tabularx}
\usepackage{array}
\usepackage{placeins}
\makeatletter
\renewcommand\subsection{\@startsection
{subsection}{2}{0mm}
 {-\baselineskip}
 {0.5\baselineskip}
{\FloatBarrier\normalfont\Large\bfseries}}
\makeatother
\newcommand{\be}{\begin{equation}}
\newcommand{\ee}{\end{equation}}

\newcommand{\PreserveBackslash}[1]{\let\temp=\\#1\let\\=\temp}

\begin{document}
\title{Ground-state phase diagram of the two-dimensional $t-J$ model}

\author{Sheng-Hao Li} \affiliation{Centre for Modern Physics and Department of Physics,
Chongqing University, Chongqing 400044, The People's Republic of
China}

\author{Qian-Qian Shi} \affiliation{Centre for Modern Physics and Department of Physics,
Chongqing University, Chongqing 400044, The People's Republic of
China}

\author{Huan-Qiang Zhou}
\affiliation{Centre for Modern Physics and Department of Physics,
Chongqing University, Chongqing 400044, The People's Republic of
China}

\begin{abstract}
The ground-state phase diagram of the two-dimensional $t-J$ model is
investigated in the context of the tensor network algorithm in terms
of the graded Projected Entangled-Pair State representation of the
ground-state wave functions. There is a line of phase separation
between the Heisenberg anti-ferromagnetic state without hole and a
hole-rich state. For both $J=0.4t$ and $J=0.8t$, a systematic
computation is performed to identify all the competing ground states
for various dopings. It is found that, besides a possible Nagaoka's
ferromagnetic state, the homogeneous regime consists of four
different phases: one phase with charge and spin density wave order
coexisting with a $p_x (p_y)-$wave superconducting state, one phase
with the symmetry mixing of $d+s-$wave superconductivity in the
spin-singlet channel and  $p_x (p_y)-$wave superconductivity in the
spin-triplet channel in the presence of an anti-ferromagnetic
background, one superconducting phase with extended $s-$wave
symmetry, and one superconducting phase with $p_x (p_y)-$wave
symmetry in a ferromagnetic background.

\end{abstract}
\pacs{74.20.-z, 02.70.-c, 71.10.Fd}
 \maketitle

Since the discovery of high temperature
superconductivity~\cite{muller}, significant progress has been made
in our understanding of strong correlation physics. It was
Anderson~\cite{anderson} who realized the importance of Mott-Hubbard
insulators and put forward the resonating valence bond picture as a
promising route towards the understanding of an electron pairing
mechanism responsible for an unprecedented high transition
temperature, which are observed for copper oxides (cuprates).
Actually, the detailed analysis of electronic states deduced from
experiments shows that the undoped parent compound is a Mott-Hubbard
insulator and the hole doping is mainly on oxygen sites, with its
effective low energy physics described by the two-dimensional $t-J$
model~\cite{zhangrice}.

Although a lot of efforts have been made to gain a full picture of the underlying physics
of the two-dimensional $t-J$ model (see, e.g.,~\cite{baskarananderson,baskaran,affleck,zhangetal,dagotto,lin,rice,poilblanc,ogata,hellberg,hellberg2,wenlee,white,sorella}),
no consensus has been achieved as to the question whether or not the
two-dimensional $t-J$ model superconducts. On the one hand, the variational Monte Carlo (VMC) method clearly indicates that
the $d_{x^2-y^2}-$wave superconductivity is stable at absolute zero temperature for a physically realistic coupling parameter
regime~\cite{tj}, and field theoretical slave-boson approximation yields qualitatively many peculiar phenomenological features of cuprate superconductors~\cite{leenagaosawen}.
On the other hand, exact diagonalization (ED) of the $t-J$ model on a small cluster and quantum Monte Carlo (QMC) simulation of the two-dimensional Hubbard model do not produce
convincing evidence supporting the existence of superconductivity in the $t-J$ model within a physically relevant regime~\cite{dagotto-t-j,dagotto}.

In this paper, we attempt to investigate the ground-state phase
diagram of the two-dimensional $t-J$ model by exploiting a
newly-developed tensor network algorithm~\cite{shi} in terms of the
graded Projected Entangled-Pair State (gPEPS) representation of the
ground-state wave functions (for an ungraded version,
see~\cite{jordan}). It is found that, the algorithm yields reliable
results for the two-dimensional $t-J$ model at and away from half
filling, with the truncation dimension up to 6. We are able to
locate a line of phase separation (PS) between the Heisenberg
anti-ferromagnetic state without hole and a hole-rich state, which
qualitatively agrees with the results based on the high-temperature
expansions (HTE)~\cite{rice}, the VMC~\cite{ogata}, and the
density-matrix renormalization method (DMRG)~\cite{white}. In the
homogeneous regime, the two-dimensional $t-J$ model exhibits very
rich physics.  Away from half filling, the regime is divided into
four different phases: one phase with charge and spin density wave
order coexisting with a $p_x (p_y)-$wave superconducting state
(CDW+SDW+PW), one phase with the symmetry mixing of $d+s-$wave
superconductivity in the spin-singlet channel and  $p_x (p_y)-$wave
superconductivity in the spin-triplet channel in the presence of an
anti-ferromagnetic background (DSW+PW+AF), one superconducting phase
with extended $s-$wave symmetry (SW), and one superconducting phase
with $p_x (p_y)-$wave symmetry in a ferromagnetic background
(PW+FM), besides a possible Nagaoka's ferromagnetic state
(FM)~\cite{shiba}.

{\it Ground-state phase diagram of the two-dimensional $t-J$ model.}
The two-dimensional $t-J$ model is described by the
Hamiltonian~\cite{dagotto} consisting of a hopping term and a super-exchange interaction:
\begin{equation}\label{ham}
H=-t \sum_{\langle ij \rangle \sigma}[{\cal P} (c^\dagger_{i \sigma}c_{j \sigma} +
{\rm H.c.}) {\cal P}]+ J \sum_{\langle ij \rangle} ({ \bf S}_i \cdot {\bf S}_j
-\frac {1}{4} n_i\;n_j),
\end{equation}
where ${ \bf S}_i$  are spin $1/2$ operators at site $i$ on a square lattice,
${\cal P}$ is the projection operator excluding double occupancy,
and $t$ and $J$ are, respectively, the hoping constant and
anti-ferromagnetic coupling between the nearest neighbor sites
$\langle ij \rangle$.

\begin{figure}
\centering
\begin{overpic}
 [width=0.45\textwidth]{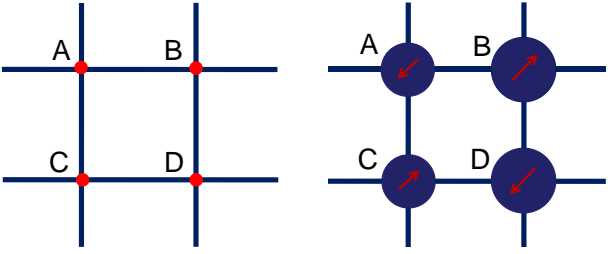}
\end{overpic}
\caption{(color online) Left panel: The unit cell of the graded Projected Entangled-Pair State representation
of the ground-state wave functions. Right panel: The pattern of the coexisting charge and spin density wave order. Here, the radius of the circles
are proportional to the fillings at sites, whereas the arrows inside the circles represent the directions and magnitudes of the spin density
wave order parameters. This represents a vertical commensurate stripe state, which breaks the four-fold rotation symmetry and the translation
symmetry in the horizontal direction for charge density wave order and in both directions for spin density wave order. } \label{orderparameter}
\end{figure}

The model is simulated by exploiting the gPEPS tensor network
algorithm~\cite{shi}, with the truncation dimension up to 6. Here,
we stress that, although the algorithm is variational in nature, an
assumption has to be made on the choice of the unit cell of the
gPEPS representation of the ground-state wave functions. For the
$t-J$ model, we have chosen the plaquette as the unit cell (see
Fig.~\ref{orderparameter}, left panel).

\begin{figure}
\centering
\begin{overpic}
  [width=0.48\textwidth]{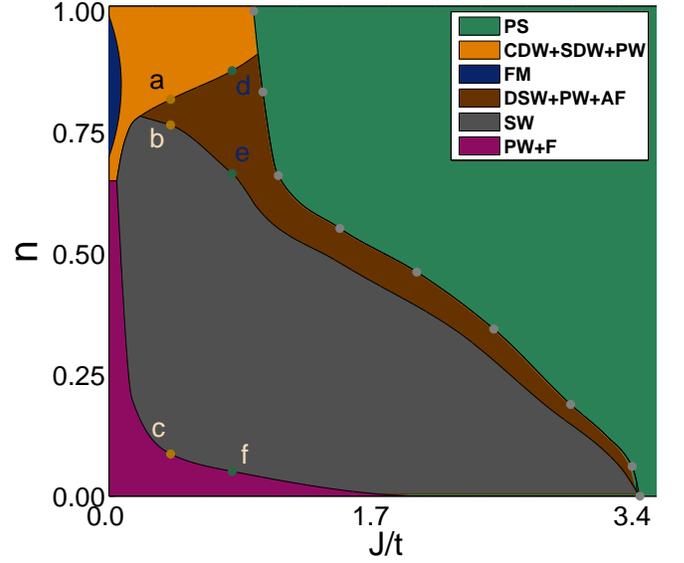}
  \end{overpic}
\caption{(color online) The proposed ground-state phase diagram of
the two-dimensional $t-J$ model.  First, for $J/t\geq 0.95$, there
is a line of phase separation (PS), whereas for $J/t\geq 0.95$, no
PS occurs. Second, the homogeneous regime is divided into four
different phases:  one phase with charge and spin density wave order
coexisting with a $p_x (p_y)-$wave superconducting state
(CDW+SDW+PW), one phase with the symmetry mixing of $d+s-$wave
superconductivity in spin-singlet channel and  $p_x (p_y)-$wave
superconductivity in spin-triplet channel in the presence of an
anti-ferromagnetic background (DSW+PW+AF), one superconducting phase
with an extended $s-$wave symmetry (SW), and one superconducting
phase with a $p-$wave symmetry (PW+F), besides a possible Nagaoka's
ferromagnetic state (FM). Here, a systematic computation has been
performed for both $J/t=0.4$ and $J/t=0.8$ with the truncation
dimension $\mathbb{D}=6$, whereas the dash lines separating
different phases are a guide for the eyes. For $J/t=0.4$, the
DSW+PW+AF phase occurs for fillings from $n=0.818$ (denoted as point
a) to $n=0.765$ (denoted as point b), the SW phase occurs for
fillings from $n=0.765$ to $n=0.087$ (denoted as point c), and the
CDW+SDW+PW and PW+FM phases occur for fillings from $n=1$ to
$n=0.818$ and from $n=0.087$ to $n=0$, respectively. For $J/t=0.8$,
the DSW+PW+AF phase occurs for fillings from $n=0.877$ (denoted as
point d) to $n=0.665$ (denoted as point e), the SW phase occurs for
fillings from $n=0.665$ to $n=0.051$ (denoted as point f), and the
CDW+SDW+PW and PW+FM phases occur for fillings from $n=1$ to
$n=0.877$ and from $n=0.051$ to $n=0$, respectively.}
\label{unitcell}
\end{figure}

In Fig.~\ref{unitcell}, we plot the ground-state phase diagram for
the two-dimensional $t-J$ model, with the vertical and horizontal
axes, $n$ and $J/t$, denoting the number of electrons per site and
the ratio between the anti-ferromagnetic coupling and the hoping
constant, respectively. At half filling, $n=1$, the $t-J$ model
reduces to the two-dimensional Heisenberg spin $1/2$ model.
Therefore, a long range anti-ferromagnetic order
exists~\cite{manou}. In this case, our algorithm yields the
ground-state energy per site, $e=-1.1675J$,  for the truncation
dimension $\mathbb{D}=4$, and $e=-1.1683J$, for the truncation
dimension $\mathbb{D}=6$. This is comparable to the best QMC
simulation result: $e = -1.1694J$~\cite{manou,huse}, although the
anti-ferromagnetic N\'eel order moment is $0.36$ versus $0.31$. Away
from half filling, the model exhibits quite different behaviors for
small and large values of the anti-ferromagnetic coupling $J$. For
$J/t \geq 0.95$, there is a line of PS between the Heisenberg
anti-ferromagnetic state without hole and a hole-rich state, whereas
for $J/t\leq 0.95$, no PS occurs. This agrees qualitatively with the
results based on the HTE~\cite{rice}, the VMC method~\cite{ogata},
and the DMRG~\cite{white}. Note that our result for the transition
point $J_c=3.45t$ at low electron density is quite close to the
exact value $J_c= 3.4367t$~\cite{hellberg}. Here, the simulation has
been performed for the truncation dimension $\mathbb{D}=6$.

We point out that a discrepancy exists concerning the PS boundary of the $t-J$ model. In Ref.~\cite{lin},  a combination of analytic and
numerical calculations is used to establish the existence of PS at all super-exchange interaction strengths. Although the simulation based on the Green's function Monte Carlo method supports this scenario~\cite{hellberg2},  many
others~\cite{rice,poilblanc,ogata,white} found that the model phase separates only for $J/t$ larger than some finite critical value around $J/t \sim 1$. That is, PS occurs {\it only}
outside the physically realistic parameter region of the $t-J$ model. With the observation that  no
significant shift with the truncation dimension $\mathbb{D}$ increasing from  $\mathbb{D}=4$~\cite{shi} to  $\mathbb{D}=6$ is found, we conclude that {\it PS does not occur for $J/t \leq 0.95$}.

In the homogeneous regime, the two-dimensional $t-J$ model exhibits
very rich physics. Away from half filling,  both the $d+s-$wave
superconductivity in the spin-singlet channel and  the $p_x
(p_y)-$wave superconductivity in the spin-triplet channel, and/or
charge and spin density wave order,  occur in different doping
regimes for a fixed $J/t$.  Here, a superconducting state is
characterized by a superconducting order parameter $\Delta \equiv
\langle \hat \Delta \rangle$, with  $\hat \Delta$  defined as
follows: For $s$-wave, $\hat \Delta_{s} = 1/(4 \sqrt{2})\;
c^{+}_{i_{x},i_{y}\uparrow}[c^{+}_{i_{x}+1,i_{y}\downarrow}+c^{+}_{i_{x}-1,i_{y}\downarrow}+c^{+}_{i_{x},i_{y}+1\downarrow}+c^{+}_{i_{x},i_{y}-1\downarrow}]
-[\uparrow \leftrightarrow \downarrow]$; for $d$-wave, $\hat
\Delta_{d} = 1/(4 \sqrt{2})\;
c^{+}_{i_{x},i_{y}\uparrow}[c^{+}_{i_{x}+1,i_{y}\downarrow}+c^{+}_{i_{x}-1,i_{y}\downarrow}-c^{+}_{i_{x},i_{y}+1\downarrow}-c^{+}_{i_{x},i_{y}-1\downarrow}]
-[\uparrow \leftrightarrow \downarrow]$; for $p_x$-wave, $\hat
\Delta_{p_x} = \hat \Delta_{p_x +} - \hat \Delta_{p_x -}$, with
$\hat \Delta_{p_x \pm} = 1/2\;
\left(c^{+}_{i_{x},i_y\uparrow}c^{+}_{i_{x}\pm1,i_y\uparrow},
[c^{+}_{i_{x},i_y\uparrow}c^{+}_{i_{x}\pm1,i_y\downarrow}+c^{+}_{i_{x},i_y\downarrow}c^{+}_{i_{x}\pm1,i_y\uparrow}]/\sqrt{2},
c^{+}_{i_{x},i_y\downarrow}c^{+}_{i_{x}\pm1,i_y\downarrow} \right)$,
and a similar definition of $\hat \Delta_{p_y}$ for $p_y$-wave.

{\it Charge and spin density wave order coexisting with a $p_x
(p_y)-$wave superconducting state.} For a physically realistic
super-exchange coupling $J$ and a hopping constant $t$ (such as
$J/t=0.4$), a spin-triplet $p_x (p_y)-$wave  superconducting state
coexists with charge and spin density wave order for dopings up to
around $\delta \sim 0.18$, with $\delta \equiv1-n$.   The occurrence
of such a spin-triplet $p_x (p_y)-$wave superconducting component is
unexpected, although this would, in our opinion, have been
anticipated from the presence of the Nagaoka's ferromagnetic
state~\cite{shiba}. In addition, Dagotto and Riera~\cite{dagotto}
observed important ferromagnetic and anti-ferromagnetic correlations
in this regime from the ED of the $t-J$ model on a small cluster,
which may be properly interpreted as the precursor of a spin-triplet
$p_x (p_y)-$wave pairing state, coexisting with charge and spin
density wave order. Note that the coexisting charge and spin density
wave order forms a pattern, as displayed in
Fig.~\ref{orderparameter} (right panel). In this phase, the
translational invariance under one-site shifts, and four-fold
rotation symmetry, as well as the $U(1)$ symmetry in the charge
sector and the $SU(2)$ symmetry in the spin sector, are
spontaneously broken.

\begin{figure}%
\centering
\begin{overpic}
 [width=0.46\textwidth,totalheight=54mm]{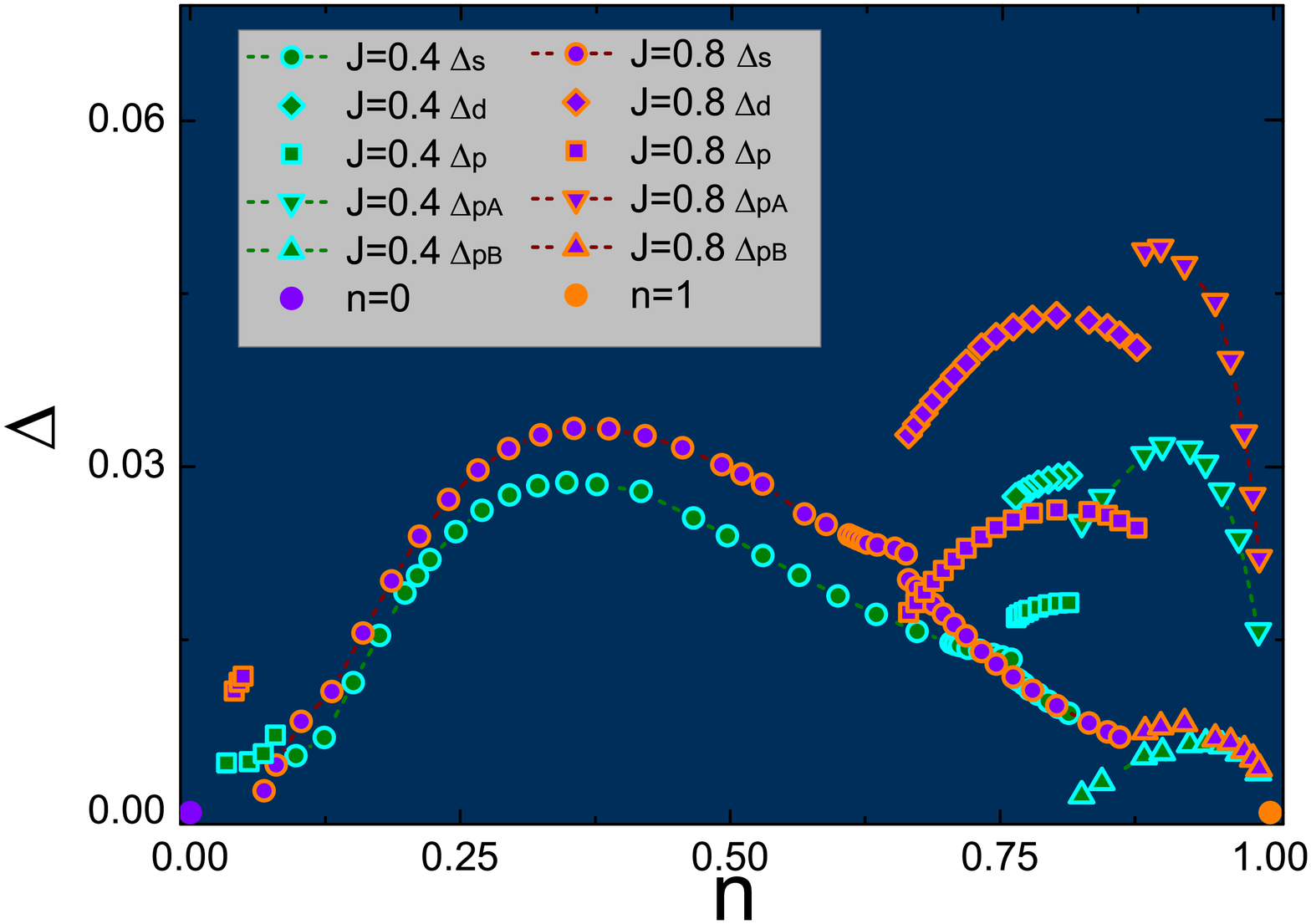}
  \end{overpic}
\begin{overpic}
[width=0.46\textwidth,totalheight=54mm]{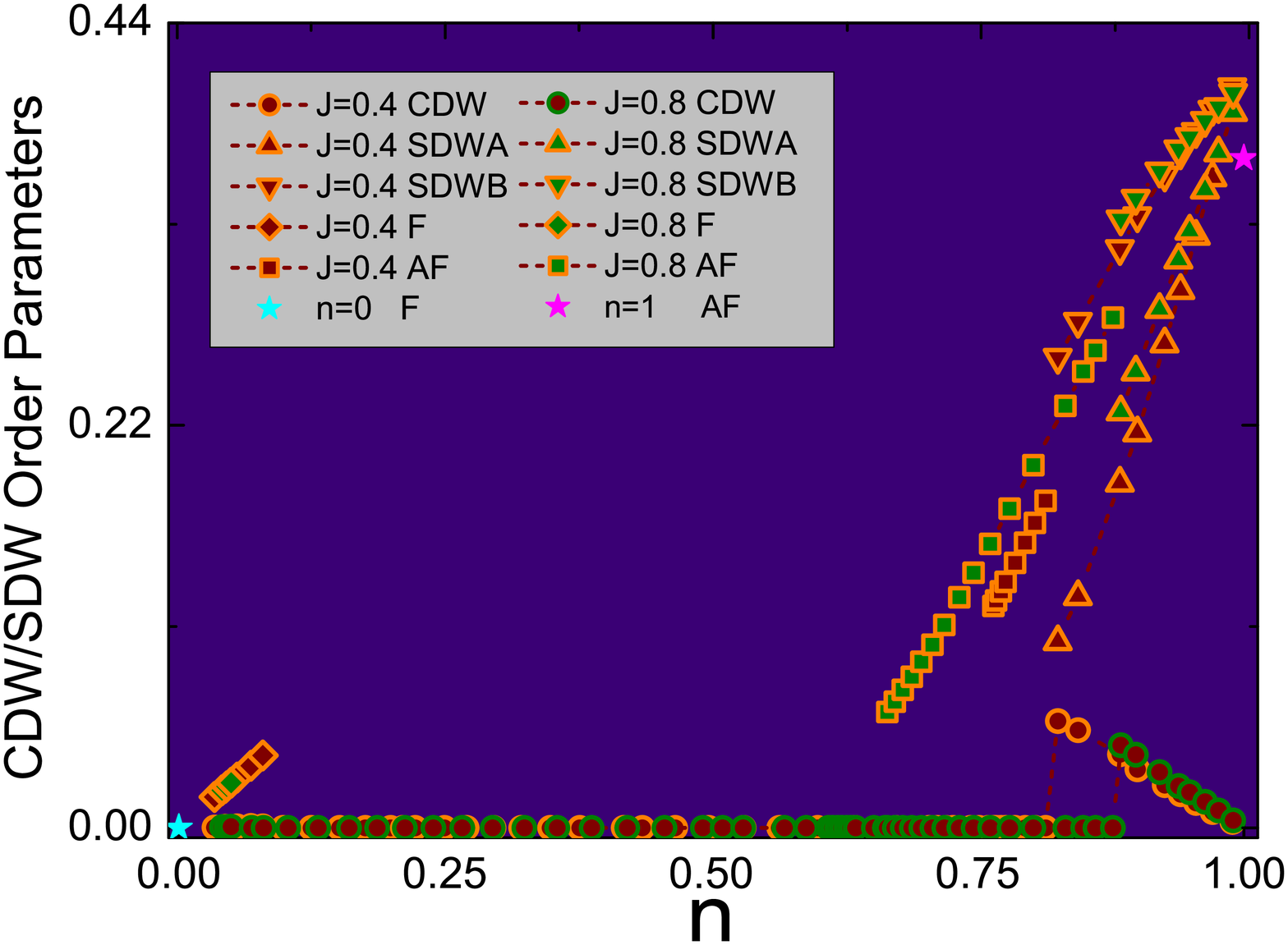}
  \end{overpic}
\caption{(color online) Upper panel: The dependence of the $d$-wave,
$s$-wave, and $p_x (p_y)-$wave superconducting order parameters, $\Delta_d$,
$\Delta_s$ and $\Delta_p$, on electron filling $n$.  Here, we
evaluate $\Delta_s$, $\Delta_d$ and $\Delta_p$ for $J/t=0.4$ and
$J/t=0.8$. Lower panel: The
dependence of the spin density wave (SDW) and the
charge density wave (CDW) order parameters in the CDW+SDW+PW phase,
the anti-ferromagnetic N\'eel order parameter in the DSW+PW+AF phase,
and the ferromagnetic order parameter in the PW+F phase on electron filling $n$.
Note that, in the CDW+SDW+PW phase, all the
order parameters are evaluated for two sites A and B in the unit
cell, with the subscripts A and B  as their labels, except for the
charge density wave order parameter that is defined as half the
difference between the densities at two sites A and B (see
Fig.~\ref{orderparameter}, right panel).
} \label{phasediagram}
\end{figure}

In Fig.~\ref{phasediagram}, the magnitude of the spin-triplet $p_x
(p_y)-$wave superconducting order parameter $\Delta_{p}$, along with
those of the charge and spin density wave order parameters, are
plotted for both $J/t=0.4$ and $J/t=0.8$, which have been evaluated
with the truncation dimension $\mathbb{D}=6$. In this phase, all the
order parameters are evaluated for two sites A and B in the unit
cell, with the subscripts A and B  as their labels, except for the
charge density wave order parameter that is defined as half the
difference between the densities at two sites A and B (see
Fig.~\ref{orderparameter}, right panel).

{\it The mixing of the spin-singlet $d+s-$wave and spin-triplet $p_x
(p_y)-$wave superconductivity in the presence of an
anti-ferromagnetic background.} As shown in Fig.~\ref{unitcell},
there is a superconducting phase with mixed spin-singlet $d+s-$wave
and spin-triplet $p_x (p_y)-$wave symmetries in the presence of an
anti-ferromagnetic background.  For $J=0.4$, it occurs for fillings
from $n=0.818$ to $n=0.765$. For $J=0.8$, it occurs for fillings
from $n=0.877$ to $n=0.665$.  In this phase, the translational
invariance under one-site shifts, and four-fold rotation symmetry,
as well as the $U(1)$ symmetry in the charge sector and the $SU(2)$
symmetry in the spin sector, are spontaneously broken. Note that the
$d$-wave, $s$-wave and  $p_x (p_y)-$wave components are homogeneous.

In Fig.~\ref{phasediagram}, the magnitudes of the $d-$wave, $s-$wave
and $p_x (p_y)-$wave order parameters, $\Delta_d$, $\Delta_s$, and
$\Delta_p$ (upper panel), along with that of the anti-ferromagnetic
order parameter (lower panel), are plotted for the $t-J$ model with
$J/t=0.4$ and $J/t=0.8$  which have been evaluated with the
truncation dimension $\mathbb{D}=6$.  One observes that the $d-$wave
component $\Delta_d$ vanishes at electron fillings $n_{c1}$:
$n_{c1}=0.765$ and $n_{c2}=0.818$ for $J/t=0.4$ and $n_{c1}=0.665$
and $n_{c2}=0.877$ for $J/t=0.8$.  We also list the numerical values
of the $d-$wave and $s-$wave components, $\Delta_d$ and $\Delta_s$,
for both $J/t =0.4$ and $J/t =0.8$ at different fillings in Table
~\ref{Tab1}, together with their ratio $\Delta_s/\Delta_d$. The fact
that the pairing symmetry (in the spin-singlet channel) is of  the
$d+s$-wave nature manifests itself in that the ratio
$\Delta_s/\Delta_d$ is always real. On the other hand, the symmetry
mixing of the spin-singlet and spin-triplet channels arises from the
spin-rotation symmetry breaking: {\it spin is not a good quantum
number}.
\begin{center}
\begin{table}
\begin{tabular}{|r|c|c|c|c|}
 \hline $J/t$ & $n$ & $\Delta_{s}$ & $\Delta_{d}$ &
$\Delta_{s}/\Delta_{d}$ \\ \cline{1-5}
      &0.768 &$0.0107 - 0.0042i$ & $-0.0256 + 0.0100i$ & -0.416\\ \cline{2-5}
      &0.776 &$0.0101 - 0.0040i$ & $-0.0261 + 0.0103i$ & -0.386\\ \cline{2-5}
      &0.785 &$0.0096 - 0.0037i$ & $-0.0266 + 0.0103i$ & -0.360\\ \cline{2-5}
 0.4  &0.794 &$0.0091 - 0.0032i$ & $-0.0270 + 0.0100i$ & -0.336\\ \cline{2-5}
      &0.804 &$0.0085 - 0.0033i$ & $-0.0270 + 0.0105i$ & -0.314\\ \cline{2-5}
      &0.814 &$0.0080 - 0.0031i$ & $-0.0272 + 0.0106i$ & -0.295\\ \cline{1-5}
      &0.720 &$0.0147 - 0.0052i$ & $-0.0375 + 0.0134i$ & -0.391\\ \cline{2-5}
      &0.722 &$0.0145 - 0.0053i$ &$-0.0378 + 0.0133i$ & -0.383\\ \cline{2-5}
      &0.725 &$0.0143 - 0.0050i$ &$-0.0382 + 0.0133i$ & -0.375\\ \cline{2-5}
      &0.741 &$0.0129 - 0.0045i$ &$-0.0398 + 0.0139i$ & -0.325\\ \cline{2-5}
 0.8  &0.757 &$0.0116 - 0.0046i$ &$-0.0401 + 0.0160i$ & -0.289\\ \cline{2-5}
      &0.800 &$0.0091 - 0.0037i$ &$-0.0419 + 0.0168i$ & -0.217\\ \cline{2-5}
      &0.811 &$0.0085 - 0.0034i$ &$-0.0421 + 0.0169i$ & -0.203\\ \cline{2-5}
      &0.833 &$0.0076 - 0.0030i$ &$-0.0422 + 0.0169i$ & -0.179\\ \cline{2-5}
      &0.842 &$0.0072 - 0.0029i$ &$-0.0419 + 0.0168i$ & -0.171\\ \cline{2-5} \hline
\end{tabular}
\caption{The numerical values of the $s$-wave and $d$-wave
superconducting order parameters, $\Delta_s$ and $\Delta_d$, and
their ratio $\Delta_s/\Delta_d$ for both $J/t =0.4$ and $J/t =0.8$
at different fillings in the DSW+PW+AF phase.}\label{Tab1}
\end{table}
\end{center}
{\it The spin-singlet superconducting phase with extended $s-$wave
symmetry.} For a fixed $0.1<J/t<0.95$, if one keeps on increasing
doping until the $d-$wave component $\Delta_d$ vanishes, then {\it
only} an $s$-wave order parameter $\Delta_s$ survives, resulting in
a spin-singlet superconducting phase with extended $s-$wave
symmetry. For $J/t=0.4$ and $J/t=0.8$, the $s-$wave pairing order
parameter $\Delta_s$ is plotted in Fig.~\ref{phasediagram}.  A
peculiar feature of the $s-$wave order parameter is that it is
almost linearly increasing with increasing doping, until it reaches
its maximum, and then it monotonically decreases.

Although this phase is perhaps only of academic interest due to its
unphysical nature of large dopings for cuprate superconductors, we
believe it is important to clarify its pairing symmetry. Note that a
small portion located between $2.0\leq J/t \leq 3.4367$ has been
identified as a spin-singlet superconducting phase with $s-$wave
symmetry in Refs.~\cite{hellberg,lin,kagan,hellberg2}.

{\it The spin-triplet superconducting phase with $p_x (p_y)-$wave
symmetry in a ferromagnetic background.} For large dopings, a
spin-triplet superconducting phase with $p_x (p_y)-$wave symmetry in
a ferromagnetic background occurs. The existence of such a
spin-triplet superconducting phase with $p_x (p_y)-$wave symmetry
has been pointed out by Kagan and Rice~\cite{kagan}. Therefore, our
simulation results are consistent with their analytical analysis.

In Fig.~\ref{phasediagram}, the magnitude of the spin-triplet
$p-$wave order parameter, together with that of the ferromagnetic
order parameter, are plotted  for both $J/t=0.4$ and $J/t=0.8$ with
the truncation dimension $\mathbb{D}=6$. Here,  the $p_x (p_y)-$wave
order parameter is identical at all sites, thus it is homogeneous.

{\it Doping-induced quantum phase transitions.} Now we turn to
doping-induced quantum phase transitions (QPTs) for the
two-dimensional $t-J$ model. For $J/t=0.4$, the model undergoes a
QPT from the CDW+SDW+PW phase to the DSW+PW+AF phase at $n=0.818$,
and a QPT from the DSW+PW+AF phase to the SW phase at $n=0.765$, and
again a QPT from the SW phase to the PW+F phase at $n=0.087$. For
$J/t=0.8$, the model undergoes a QPT from the CDW+SDW+PW phase to
the DSW+PW+AF phase at $n=0.877$, and a QPT from the DSW+PW+AF phase
to the SW phase at $n=0.665$, and again a QPT from the SW phase to
the PW+F phase at $n=0.051$. In addition,  a QPT occurs between the
DSW+PW+AF phase and the PW+F phase. Here, all QPTs are first-order,
as one may judge from the behaviors of the order parameters shown in
Fig.~\ref{phasediagram}. In passing, we point out that the above
results have been deduced from the ground-state fidelity approach in
the context of tensor network representations~\cite{zhou1}, with
superconducting order parameters being read off from both one-site
and two-site reduced density matrices according to a general scheme
advocated in Ref.~\cite{zhou2}.

In summary, we have investigated the ground-state phase diagram of
the two-dimensional $t-J$ model in the context of the tensor network
algorithm. The relevance of our simulation results to the high
temperature cuprate superconductors will be discussed in a
forthcoming publication~\cite{zhou-htsc}.

This work is supported in part by the National Natural Science
Foundation of China (Grant Nos: 10774197 and 10874252).

\end{document}